\documentclass[11pt]{article}
\usepackage{a4wide}                      
\usepackage{epsf}                        
\usepackage{latexsym}                    
\usepackage{amsfonts}                    
\usepackage{amssymb}                     

\newcommand{\sect}[1]{\setcounter{equation}{0}\section{#1}}
\renewcommand{\theequation}{\arabic{section}.\arabic{equation}}

\def\be{\begin{equation}}
\def\ee{\end{equation}}
\def\bea{\begin{eqnarray}}
\def\eea{\end{eqnarray}}

\def\part{\partial}

\def\incl{\mbox{i}}

\def\R{\ensuremath{\mathbb{R}}}

\def\makeatletter{\catcode`\@=11}
\makeatletter
\def\mathbox#1{\hbox{$\m@th#1$}}%
\def\math@ccstyles#1#2#3#4#5#6#7{{\leavevmode
      \setbox0\mathbox{#6#7}%
      \setbox2\mathbox{#4#5}%
      \dimen@ #3%
      \baselineskip\z@\lineskiplimit#1\lineskip\z@
      \vbox{\ialign{##\crcr
             \hfil \kern #2\box2 \hfil\crcr
             \noalign{\kern\dimen@}%
             \hfil\box0\hfil\crcr}}}}
\def\mathaccstyles{\math@ccstyles\maxdimen}
\def\maththroughstyles{\math@ccstyles{-\maxdimen}}
\def\unity%
 {\maththroughstyles{.45\ht0}\z@\displaystyle {\mathchar"006C}\displaystyle 1}
\begin{document}

\rightline{FFUOV-05/02}
\rightline{hep-th/0505073}
\rightline{May 2005}
\vspace{3.5cm}

\centerline{\LARGE \bf Fuzzy 5-spheres and pp-wave Matrix Actions}
\vspace{1.5truecm}

\centerline{
    {\large \bf Yolanda Lozano}\footnote{E-mail address:
                                  {\tt yolanda@string1.ciencias.uniovi.es}}
    {\bf and}
    {\large \bf Diego Rodr\'{\i}guez-G\'omez}\footnote{E-mail address:
                                  {\tt diego@fisi35.ciencias.uniovi.es}}
                                                            }

\vspace{.4cm}
\centerline{{\it Departamento de F{\'\i}sica,  Universidad de Oviedo,}}
\centerline{{\it Avda.~Calvo Sotelo 18, 33007 Oviedo, Spain}}

\vspace{2.5truecm}

\centerline{\bf ABSTRACT}
\vspace{.5truecm}

\noindent Using the action describing $N$ coincident gravitational waves in M-theory
we construct a pp-wave Matrix model containing a fuzzy 5-sphere giant graviton
solution. This fuzzy 5-sphere 
is constructed as a $U(1)$ fibration
over a fuzzy $CP^2$, and has the correct dependence of the radius with the 
light-cone momentum,
$r^4\sim p^+$, to approach the 5-sphere giant graviton solution of Mc.Greevy et al in the
large $N$ limit.


\newpage
\section{Introduction}

The BMN Matrix model \cite{BMN} is an example of Matrix theory
in a non-trivial background, the maximally supersymmetric pp-wave background
of M-theory \cite{FP}. Compared to Matrix theory in a flat background it contains 
additional mass and dielectric terms that remove the flat directions \cite{BMN,DSR},
leaving isolated vacua in the form of giant
gravitons \cite{GST}.  String theory in this background is solvable \cite{Metsaev},
and through the AdS/CFT correspondence, Type IIB string theory in the pp-wave
background has been shown to be
dual to the, so-called, BMN sector of four dimensional ${\cal N}=4$ SYM \cite{BMN}.

The derivation of the BMN Matrix model was based on the generalization to arbitrary number
of particles of the action for
a superparticle in the pp-wave background, under the requirement of
consistency with supersymmetry \cite{BMN}. 
The same Matrix model was obtained in \cite{DSR}
by regularizing the light-cone supermembrane action in the pp-wave
background, in the same way Matrix theory in a flat background arises from the flat
supermembrane of \cite{WHN} (see also \cite{SY1}).
Moreover, the model was shown to be related to the action
for Type IIA D0-branes upon dimensional reduction plus Sen-Seiberg \cite{S1,S2} limit.

{}From the last point of view one could say that the basic degrees of freedom of
the BMN Matrix model are gravitational waves propagating in the light-cone direction.
An action describing coincident gravitational waves in arbitrary M-theory
backgrounds has been constructed in \cite{JL2}. This action 
goes beyond the linear order approximation of Matrix theory in a weakly curved background
\cite{KT}
precisely by imposing consistency with the action for coincident Type IIA D0-branes upon
reduction along the direction of propagation. One non-trivial check of the validity of this action
is that it has been successfully used in the
study of gravitational waves in $AdS_m \times S^n$ backgrounds,
which are not linear perturbations to Minkowski, and in particular in the microscopical
description of giant gravitons  in terms of dielectric
gravitational waves\footnote{Of course, to describe these giant gravitons in non M-theory AdS 
type backgrounds the analogous
action for coincident waves has to be constructed in Type II, but this can easily
be done using dualities from the M-theory action.} 
\cite{JL2,JLR,JLR2,JLR3}.

For consistency the BMN Matrix model should come up as the action of \cite{JL2} in the maximally
supersymmetric pp-wave background of M-theory. Indeed, one can check
that when the waves propagate in this background with
a non-vanishing light-cone momentum the action
of \cite{JL2} reduces to the BMN Matrix model. In this derivation the dielectric coupling present in 
the BMN Matrix model simply arises as the non-Abelian Myers coupling \cite{myers}
to the 3-form potential
in the action for the waves. As shown in \cite{BMN} this coupling
supports a zero-energy solution consisting on the expansion of the waves by dielectric effect
into a fuzzy 2-sphere with light-cone momentum, which
constitutes the microscopical version of the 2-sphere giant graviton solution of \cite{GST}
in this background.

A 5-sphere giant graviton solution is also known in the M-theory pp-wave background \cite{GST},
a microscopical description to which has not successfully been given so far.
One would expect that microscopically the waves would expand into a fuzzy 5-sphere through their
(quadrupolar)
coupling to the 6-form potential of the background\footnote{Since a 5-sphere has 5 relative
dimensions with respect to a point-like object, the 6-form potential has to be contracted as
well with some other (Abelian) direction of the background in order to be able to couple to 
a one dimensional worldvolume, as we will see.}. However, such a coupling does not exist
in the BMN Matrix model, this being the reason why a fuzzy 5-sphere giant graviton 
solution cannot easily be constructed. This is of course related to the difficulty in seeing the 
M5-brane in the Matrix model \cite{BSS}.

As one would expect, this problem is already encountered in the $AdS_4\times S^7$ and
$AdS_7\times S^4$ backgrounds from which the M-theory pp-wave emerges after
Penrose limit. In these backgrounds there is a 5-sphere giant graviton solution which expands
either into the $S^7$ or the $AdS_7$ parts of the geometry\footnote{The giant
graviton expanding into the anti-deSitter part of the geometry is referred in the literature as
dual giant graviton \cite{GMT,HHI}.} \cite{GST,GMT}. 
The microscopical description of these giant gravitons
in terms of dielectric gravitational waves has been given recently in \cite{JLR3}. 
In this reference it is shown that in order to find the quadrupolar couplings to the 6-form
potential responsible for the expansion of the waves it is necessary to go to a more non-perturbative
regime of M-theory. This regime is reached by interchanging the direction of propagation of
the waves with another compact direction of the background with a geometrical meaning, 
that we will clarify below. Reduction along the direction of propagation gives rise to Type IIA
D0-branes, which upon T-duality along the 
geometrical compact direction that they inherit from M-theory
give rise to Type IIB
D-strings. On the other hand, reduction of the waves along the geometrical 
compact direction gives rise to
Type IIA waves, which upon 
T-duality along their direction of propagation give rise to Type IIB F-strings. 
The S-duality between the F-strings and the D-strings is therefore generated in 
M-theory through the interchange of the direction of propagation and the geometrical
compact direction.

{}From this point of view
the action of \cite{JL2} describes {\it perturbative} gravitational waves. The details of this action
(see \cite{JL2} and \cite{JL1})
show that the direction of propagation of the waves is isometric, being in this sense special.
The discussion in the previous paragraph
suggests that we can use
the same action to describe {\it non-perturbative} gravitational waves if the isometric direction is
instead identified with some compact geometrical direction of the background.
Indeed, one can see that doing this in the $AdS_4\times S^7$ and $AdS_7\times S^4$
backgrounds one finds non-vanishing
quadrupolar couplings to the 6-form
potential that can cause the expansion of the waves into fuzzy 5-spheres \cite{JLR3}.
These fuzzy 5-spheres are
defined as $S^1$ bundles over fuzzy $CP^2$, and the direction along the $S^1$ is the one
playing the
role of compact geometrical direction. The strongest check
of the validity of this action is that the corresponding microscopical giant and dual giant graviton solutions 
have the same radii and satisfy the same bound (for the giant graviton), for
large number of gravitons, than the classical 5-sphere solutions of \cite{GST} and \cite{GMT}.

The expansion of the
waves into fuzzy 5-spheres is from the previous point of view more non-perturbative than the 
corresponding expansion into fuzzy 2-spheres.  This is in agreement with
previous observations along these lines \cite{BMN,MSR}.
In these references it is argued that the 5-brane cannot appear as a classical solution to
the BMN Matrix model because the scaling of its radius with the coupling constant is more
non-perturbative than the one corresponding to a classical solution, and it is compared to the
difficulty in realizing classically the process in which $N$ D3-branes blow up into an NS5-brane
found in \cite{PS} in mass deformed ${\cal N}=4$ SYM. In this case a non-perturbative action for 
coincident D3-branes would show a dielectric coupling to the NS-NS 6-form potential that would
cause the expansion of the D3-branes into an NS5-brane.
Our discussion above
shows that this difficulty is also present in M-theory, and moreover, that it is not peculiar 
to the pp-wave background.

In this paper we will use the action of \cite{JL2} to construct a {\it non-perturbative} Matrix
model which is solved by a non-commutative 5-sphere. We will see that this 5-sphere 
has the same radius and energy than the classical 5-sphere giant graviton solution 
of \cite{GST} in this background. We will also comment in the conclusions 
on the relation between our solution
and previous attempts in the literature to obtain the fuzzy 5-sphere giant graviton as a 
supersymmetry preserving solution \cite{Ram2,S-J,Nastase}.

 \section{The action for M-theory gravitational waves}
 
 The worldvolume theory associated to $N$ coincident gravitational waves in M-theory
 is a $U(N)$
 gauge theory, in which the vector
 field is associated to M2-branes (wrapped on the direction of propagation of the waves)
 ending on them \cite{JL2}. This vector field gives the BI field living in a set of coincident 
 D0-branes upon reduction along the direction of propagation of the
 waves. 
 
 In this paper we will use a truncated version of the action in \cite{JL2}
 in which the vector field is set to zero, given that it will not play any role in the backgrounds
 that we will be discussing. This action is given by 
 $S=S^{\rm BI}+S^{\rm CS}$, with
 \begin{equation}
 \label{BIaction}
 S^{\rm BI}=-\int dt\, {\rm STr} \{ k^{-1}\sqrt{-P[E_{00}+E_{0i}
 (Q^{-1}-\delta)^i_k E^{kj}E_{j0}]{\rm det Q}}\}\, ,
 \end{equation}
 where
 \begin{eqnarray}
 \label{mred}
 &&E_{\mu\nu}={\cal G}_{\mu\nu}+k^{-1}(\incl_k C^{(3)})_{\mu\nu}\,, \qquad
 {\cal G}_{\mu\nu}=g_{\mu\nu}-\frac{k_\mu k_\nu}{k^2} \\
 &&Q^i_j=\delta^i_j+ik[X^i,X^k]E_{kj}\, , \nonumber
 \end{eqnarray}
and
 \begin{equation}
 \label{CSaction}
 S^{\rm CS}= \int dt\, {\rm STr} \{ -P[k^{-2} k^{(1)}]+iP[(\incl_X \incl_X)C^{(3)}] +
 \frac12 P[(\incl_X\incl_X)^2\incl_k C^{(6)}] +\dots\}\, ,
 \end{equation}
 where the dots include couplings to higher order background potentials and products of
 different background fields contracted with the non-Abelian scalars\footnote{These 
 couplings are
 not shown explicitly because they will not play a role in the backgrounds under consideration
 in this paper.}. We have also taken $T_0$, the tension of a single gravitational wave, equal
to one. 
In this action
 $k^\mu$ is an Abelian Killing vector which, by construction,
 points on the direction of propagation of the
 waves. This direction is isometric, because the background fields are
 either contracted with the Killing vector, so that any component along the isometric
 direction of the contracted field vanishes, or pulled back in the worldvolume with
 covariant derivatives relative to the isometry (see \cite{JL2} for their explicit 
 definition)\footnote{The reduced metric ${\cal G}_{\mu\nu}$ appearing in (\ref{mred})
 is in fact defined such that its pull-back with ordinary derivatives equals the pull-back
 of $g_{\mu\nu}$ with these covariant derivatives.}.

We recall very briefly from \cite{JL2} that the action (\ref{BIaction}) $+$ (\ref{CSaction})
was obtained by uplifting to eleven dimensions the action for Type 
IIA gravitational waves derived  in \cite{JL1} using Matrix String Theory in a weakly curved background, 
and then going beyond the weakly curved background approximation
 by demanding agreement with Myers action for D0-branes 
when the waves propagate along the eleventh direction. 
In the action for Type IIA waves the 
circle in which Matrix theory is compactified in order to construct Matrix String theory
cannot be 
decompactified in the non-Abelian case \cite{JL1}. In fact, the action exhibits a 
$U(1)$ isometry associated to translations along this direction, which by construction is
also the direction on which the waves propagate.
A simple way to see this is to recall that the 
last operation in the 9-11 flip involved in the construction of Matrix String theory
is a T-duality from fundamental strings wound around the 
9th direction. Accordingly, in the action we find a minimal coupling to 
$g_{\mu 9}/g_{99}$ which is the momentum operator $k_\mu/k^2$ 
if $k^\mu=\delta^\mu_9$. Therefore, by 
construction, the action is designed to describe BPS waves with 
momentum charge along the compact isometric direction.
 It is important to mention that in the Abelian limit, when all dielectric couplings and $U(N)$ covariant
 derivatives\footnote{Which are of course implicit in the pull-backs of the action.}
 disappear, the action can be Legendre transformed into one in
 which the dependence on the isometric direction has been restored. This action is
 precisely the usual one for a massless particle written in terms of an auxiliary $\gamma$  
 metric (see \cite{JL2} and \cite{JL1} for the details), where no information remains about the 
 momentum charge carried by the particle.  
 
 As we have mentioned in the introduction the action (\ref{BIaction}) $+$ (\ref{CSaction})
 has been successfully used in the
 microscopical study of giant graviton configurations in backgrounds which are not linear
 perturbations to Minkowski, like the M-theory backgrounds
 $AdS_4\times S^7$ and $AdS_7\times S^4$ \cite{JL2,JLR3}.
In all cases perfect agreement with the
 description of \cite{GST,GMT} has been found in the limit of large number of gravitons, in
 which the commutative configurations of \cite{GST,GMT} become an increasingly better
 approximation to the non-commutative microscopical configurations \cite{myers}.
 
 In the next section we will use the same action 
 to describe gravitational waves propagating in
 the maximally supersymmetric pp-wave background of M-theory. 
 In order to find the quadrupolar coupling to the 6-form potential of this
 background we will need to interchange the direction of propagation of the waves
 (the Killing direction in the action) with a compact direction that has to do with
 the $U(1)$ decomposition of the 5-sphere contained in the background as an $S^1$ bundle over
 the two dimensional complex projective space, $CP^2$.
 
 \section{The BMN Matrix model with coupling to the 6-form potential}
 
 Let us start by recalling the form of the maximally supersymmetric pp-wave background of 
 M-theory \cite{KG,FP}.  Starting with the $AdS_4\times S^7$ background
 \footnote{One can also start from the $AdS_7\times S^4$ background
 and perform a similar limit.}, 
 written as:
 \begin{eqnarray}
 &&ds^2=L^2(-\cosh^2{\rho}\,d\tau^2+d\rho^2+\sinh^2{\rho}\,d\Omega_2^2)
 +4L^2(d\theta^2+\cos^2{\theta}d\phi^2+\sin^2{\theta}d\Omega_5^2)\, , \nonumber\\
 \nonumber\\
 &&C^{(3)}_{\tau \alpha_1\alpha_2}=-L^3\sinh^3{\rho}\sqrt{g_\alpha}\, , \qquad
 C^{(6)}_{\phi\gamma_1\dots\gamma_5}=-(2L)^6\sin^6{\theta}\sqrt{g_\gamma}\, ;
 \end{eqnarray}
 where $L$ is the radius of curvature of $AdS_4$,
$\{\alpha_i\}$ ($\{\gamma_i\}$) are the angle variables parametrizing the 2-sphere
 (5-sphere) in the notation
\begin{equation}
d\Omega_n^2=d\beta_1^2+\sin^2{\beta_1}(d\beta_2^2+\sin^2{\beta}_2(\dots +
\sin^2{\beta_{n-1}}d\beta_n^2))\, ,
\end{equation}
and $\sqrt{g_\beta}$ is the volume element on the unit $n$-sphere;
and defining  
 \begin{equation}
 x^+=\frac{3}{2\mu}(\tau+2\phi)\, , \qquad x^-=\frac{\mu L^2}{6}(\tau-2\phi)\, , \qquad
 \rho=\frac{r}{L}\, ,\qquad \theta=\frac{y}{2L}\, ,
 \end{equation}
 the M-theory maximally supersymmetric pp-wave background 
 is obtained  in the limit $L\rightarrow\infty$ \cite{BFHP,BMN}:
 \begin{eqnarray}
 \label{ppwave}
 &&ds^2=-4 dx^+ dx^- -\Bigl[(\frac{\mu}{3})^2r^2+(\frac{\mu}{6})^2y^2\Bigr] 
 (dx^+)^2+dr^2+
 r^2d\Omega_2^2+dy^2+y^2d\Omega_5^2= \nonumber\\
 \nonumber\\
 &&=-4dx^+dx^--\Bigl[(\frac{\mu}{3})^2 (x_1^2+x_2^2+x_3^2)+
 (\frac{\mu}{6})^2 y^2 \Bigr](dx^+)^2+dx_1^2+dx_2^2+dx_3^2+dy^2+y^2d\Omega_5^2
\, , \nonumber\\
 \nonumber\\
 &&C^{(3)}_{+ij}=\frac{\mu}{3}\epsilon_{ijk}x^k\, ,\quad i,j=1,2,3\, ,\qquad
 C^{(6)}_{+\gamma_1\dots\gamma_5}=-\frac{\mu}{6}y^6\sqrt{g_\gamma}\, ,
 \end{eqnarray}
where $(x_1,x_2,x_3)$ parametrize a point in $\R^3$ \footnote{We have not chosen
Cartesian coordinates in the $\R^6$ part, as in \cite{BMN}, because for our purposes
it will be more convenient to describe the 5-sphere as an $S^1$ bundle over $CP^2$,
as we discuss below.}.
 
 The BMN Matrix model gives the dynamics of DLCQ M-theory 
 in this background along the direction $x^-\sim x^-+2\pi R$, in the
 sector with momentum $2p^+=-p_-=N/R$ \cite{BMN}.
 It is a $U(N)$ Matrix theory sum of the usual
 Matrix theory of \cite{BFSS}, a term adding mass to the scalar and fermion fields and a 
 dielectric coupling to the 3-form potential.
 In this section
 we are going to show that the same Matrix
 model, plus a coupling to the six-form potential, arises from the actions (\ref{BIaction})
 and (\ref{CSaction}) when the waves propagate in the background (\ref{ppwave})
 along the $x^-$ direction.
 
As we have mentioned above,
 to do this it is convenient to describe the 5-sphere 
 as an $S^1$ bundle over $CP^2$, and introduce adapted coordinates to the
 $U(1)$ isometry associated to the $S^1$:
 \begin{equation}
 \label{adapted}
 d\Omega_5^2=(d\chi-A)^2+ds_{CP^2}^2\, ,
 \end{equation}
 with $A$ the connection providing the necessary twist in the fibre to obtain the $S^5$
 as the global space. 
 
 $CP^2$ is the coset manifold $SU(3)/U(2)$, and it is most conveniently described for 
 our purposes
 as a submanifold of $\R^8$, determined by a set of four independent constraints
 (see for instance \cite{ABIY}):
 \begin{eqnarray}
 \label{constr}
 &&\sum_{a=1}^{8}z_a z_a =1\nonumber\\
 &&\sum_{b,c=1}^8 d^{abc}z_b z_c=\frac{1}{\sqrt{3}}z_a\, ,
 \end{eqnarray}
 where $\{z_1,\dots ,z_8\}$ parametrize a point in $\R^8$, and
 $d^{abc}$ are the components of the totally symmetric $SU(3)$-invariant tensor
 defined by
 \begin{equation}
 \lambda^a \lambda^b=\frac23 \delta^{ab} + (d^{abc} + if^{abc})\lambda^c\, ,
 \end{equation}
 where $\lambda^a$, $a=1,\dots ,8$ are the Gell-Mann matrices.
 
 As we have discussed in the introduction we can construct a {\it non-perturbative} 
action for M-theory
 waves by taking $k^\mu=\delta^\mu_\chi$ in the actions (\ref{BIaction}), (\ref{CSaction}),
 with $\chi$ the coordinate adapted to the $U(1)$ fibre in (\ref{adapted}).
 Using this coordinate and the Cartesian coordinates $\{z_1,\dots ,z_8\}$ 
 embedding the $CP^2$ in $\R^8$
 the background metric and potentials read
 \begin{eqnarray}
 \label{backadap}
 &&ds^2=-4dx^+dx^- -\Bigl[(\frac{\mu}{3})^2(x_1^2+x_2^2+x_3^2)+
 (\frac{\mu}{6})^2 y^2\Bigr]
 (dx^+)^2 \nonumber\\
 \nonumber\\
 &&\hspace{0.7cm}+dx_1^2+dx_2^2+dx_3^2+dy^2+y^2[(d\chi-A)^2+dz_1^2+\dots +dz_8^2]\, ,
 \nonumber\\
 \nonumber\\
 &&C^{(3)}_{+ij}=\frac{\mu}{3}\epsilon_{ijk}x^k\, , \qquad i,j=1,2,3\, ,\nonumber\\
 \nonumber\\
 &&C^{(6)}_{+\chi abcd}=\frac{\mu}{3}y^6 f^{[abe} f^{cd]f} z^e z^f\, ,
 \qquad a,b,c,d=1,\dots 8\, ,
 \end{eqnarray}
 where $f^{abc}$ are the structure constants of $SU(3)$.  
 Note that the choice of adapted coordinates to the $U(1)$ isometry in the
 decomposition of the 5-sphere as an $S^1$ bundle over $CP^2$ has reduced the 
 explicit invariance
 of the 5-sphere from $SO(6)$ to $SU(3)\times U(1)$ \footnote{The whole invariance under 
 $SO(6)$ should however still be present in a non-manifest way.}. Therefore the
 background is manifestly invariant under $U(1)^2\times SO(3)\times SU(3)\times U(1)$,
 where the first $U(1)^2$ is associated to the translations along the $x^-$ and
 $x^+$ directions. The second direction will be identified with the worldline time, taking
 light-cone gauge $x^+=t$. The first direction is in turn taken as the direction of propagation
 of the waves, and will be a commutative direction in the action. Also, $y$, the radius of the
 5-sphere, is taken to be commutative, consistently with the invariance of the background.

 It is straightforward to substitute in the CS action (\ref{CSaction}). One finds
 \begin{equation}
 S^{\rm CS}=-\frac{\mu}{3}\int dx^+ {\rm STr} \Bigl\{ -i\epsilon_{ijk}X^k X^j X^i+
 \frac12 y^6 f_{[abe}f_{cd]f} Z^d Z^c Z^b Z^a Z^e Z^f\Bigr\}\, ,
 \end{equation}
where we have denoted with capital letters the non-commutative transverse scalars. 
 Similarly, one finds for the BI part:
 \begin{equation}
 \label{BIaction2}
 S^{\rm BI}=-\int dx^+ {\rm STr}
 \Bigl\{\frac{1}{y}\sqrt{\beta+4\dot{x}^- -\dot{X}^2-\dot{y}^2-
 y^2\dot{Z}^2}\,\Bigl(\unity-\frac{y^2}{4}[X,X]^2-\frac{y^6}{4}[Z,Z]^2\Bigr)
 \Bigr\}\, ,
 \end{equation}
 where, in our notation $[X,X]^2\equiv [X^i,X^j][X^i,X^j]$,
 $[Z,Z]^2\equiv [Z^a,Z^b][Z^a,Z^b]$, 
 \begin{equation} 
 \beta=[(\frac{\mu}{3})^2(X_1^2+X_2^2+X_3^2)+(\frac{\mu}{6})^2 y^2]
 \end{equation}
 and $\unity-\frac{y^2}{4}[X,X]^2-\frac{y^6}{4}[Z,Z]^2$ arises as the 
 expansion of the square
 root of the determinant of $Q$ up to fourth order in the embedding scalars\footnote{The
 same approximation is taken inside the square root in (\ref{BIaction2}). Note that
 this is the usual approximation taken in non-Abelian BI actions \cite{myers},
 which is valid when the non-Abelian action is good
 to describe the system of waves, that is, when the waves are distances away less than
 the Planck length (in our units $l_p=(\sqrt{2\pi})^{-1}$.},  
 for $Q$ given by:
 \begin{eqnarray}
 &&Q^i_j=\delta^i_j+iy[X^i,X^j]\, , \qquad i,j=1,2,3 \nonumber\\
 &&Q^a_b=\delta^a_b+iy^3[Z^a,Z^b]\, , \qquad a,b=1,\dots ,8  \nonumber\\
 &&Q^i_a=Q^a_i=0 \qquad \forall\, i,a\, .
 \end{eqnarray}
 In order to obtain this expression we have taken the commutators
 $[X^i,Z^a]=0$ for all $i,a$, consistently with
 the $SO(3)\times SU(3)\times U(1)$ invariance of the background (\ref{backadap}).
 In fact,
 the most general non-commutative ansatz compatible with this symmetry is to take
 \begin{equation}
 \label{ansatz1}
 X^i=\frac{r}{\sqrt{C_N}}J^i\, , \qquad i=1,2,3\, ,
 \end{equation}
 where the $J^i$ form an $N\times N$ representation of $SU(2)$ (in our conventions
 $[J^i,J^j]=2i\epsilon^{ijk}J^k$) and $C_N$ is the quadratic Casimir  in this
 representation, and
 \begin{equation}
 \label{ansatz2}
 Z^a=\frac{1}{\sqrt{C_N}}T^a\, , \qquad a=1,\dots ,8\, ,
 \end{equation}
 where the $T^a$ form an $N\times N$ representation of $SU(3)$ (in our conventions
 $[T^a,T^b]=if^{abc} T^c$), and $C_N$ is the quadratic Casimir of $SU(3)$ in this
 representation. With this ansatz:
 \begin{equation}
 \sum_{i=1}^3(X^i)^2=r^2\unity\, , \qquad {\rm and} \qquad 
 \sum_{a=1}^8(Z^a)^2=\unity\, ,
 \end{equation}
 which are the non-commutative analogues of $\sum_{i=1}^3 (x^i)^2=r^2$, which is
 satisfied by $(x_1,x_2,x_3)$ in (\ref{ppwave}), and the first constraint
 in (\ref{constr}) respectively\footnote{We will see in the next section that in order to 
 fulfil all the constraints (\ref{constr})
 we will have to take the $Z^a$ in specific $N\times N$ representations.}.
 Then $\beta$, $\dot{X}^2$ and $\dot{Z}^2$ are multiples 
 of the identity matrix. We should stress however that this is not the most general
 non-commutative ansatz one could consider.
 
 Since we are interested in the sector of the theory with a fixed value of the 
 light-cone momentum $p_-$ it is most adequate to 
 perform a Legendre transformation from $\dot{x}^-$ to $p_-$.
 Having all terms inside the square root in (\ref{BIaction2}) as multiples of the
 identity matrix, as with the ans\"atze above, simplifies a lot this transformation, 
 since the square root can be taken out of the symmetrized trace. However,
  before we make a clear statement about the non-commutativity that the
  embedding scalars should satisfy 
 it is important to recall that the action (\ref{BIaction}) $+$
 (\ref{CSaction}) describes waves which, by construction, carry $N$ units of momentum
 charge in the $\chi$ direction.
 Therefore, this momentum charge will have to
 be set to zero if we want to describe the sector of the theory with, only,
 light-cone momentum. 
 
 The difference between $p_\chi$ being zero or not is merely a
 coordinate transformation, a boost in $\chi$. However, how to perform coordinate
 transformations in non-Abelian actions is an open problem \cite{BS,H,BSW,BFLR}.
 This
 problem was already encountered in reference \cite{JLR3},
 in the microscopical study of 5-sphere giant gravitons in terms of dielectric gravitational waves.
 A careful study on how this
 limit should be taken can be found there. An essential ingredient is the
 comparison with the complementary description of the 5-sphere giant graviton configurations in
 terms of classical M5-branes, and an important check of the validity of the prescription
 given there
 is the exact agreement between the two descriptions in the limit of large number of gravitons.
 
 The basic recipe that is derived
 from the analysis in \cite{JLR3} is that only those terms that remain finite when 
 $N\rightarrow 0$ must be kept in the Hamiltonian. However, note that to obtain this limit one
 has to know 
 the non-commutativity that the embedding scalars satisfy, since there are powers
 of $N$ implicit in the quadratic Casimirs of the groups involved in the non-commutative
 definitions (see for instance (\ref{ansatz1}) and (\ref{ansatz2})). 
 Therefore, to continue further it is necessary to specify the non-commutativity that the
 transverse scalars satisfy. 
 
 Taking the 
 most general ansatz compatible with the symmetry of the background, namely
  (\ref{ansatz1}) and (\ref{ansatz2}), one obtains the  Hamiltonian 
   \begin{eqnarray}
 \label{newaction}
 H&=&-\int dx^+ {\rm STr}\Bigl\{ \frac{1}{4R}(\dot{X}^2+\dot{y}^2+y^2
 \dot{Z}^2)-\frac{1}{4R}(\frac{\mu^2}{9}X^2+\frac{\mu^2}{36}y^2)
 +\frac12 R [X,X]^2\nonumber\\
 &&-\frac{1}{16} R \,y^{10} [Z,Z]^4+i\frac{\mu}{3}
 \epsilon_{ijk}X^kX^jX^i-\frac{\mu}{6}y^6 f_{[abe}f_{cd]f} Z^d Z^cZ^b
 Z^aZ^e Z^f \Bigr\}\, ,
 \end{eqnarray}
 where in order to arrive at this expression one has to make use of expressions 
 (\ref{rep1}) and (\ref{rep2}), which are included in section 4, where we discuss in
 more detail the non-commutativity of the fuzzy 5-sphere.

 Comparing with the BMN Matrix model we find the same expression in the
$\R^3$ part of the geometry, where we are using the same Cartesian coordinates 
 $(X_1,X_2,X_3)$.
 Regarding the $\R^6$ part, BMN use Cartesian coordinates,
 in which the $SO(6)$ invariance
 is manifest,
  whereas we are using coordinates
 in which this invariance is reduced to $SU(3)\times U(1)$, and, moreover, we have
 implicitly taken 
 a particular non-commutative ansatz for these coordinates. It could be that, apart from 
 the new dielectric coupling involving the $Z^a$ coordinates, both actions were
 equivalent through the right coordinate transformation, although we have not
 checked this out explicitly. In any case a non-trivial check of the validity of our 
 Hamiltonian (\ref{newaction}) is that we will find a new fuzzy 5-sphere
 solution with the correct radius in the large $N$ limit.
 
 In the next section we will see how the fuzzy 5-sphere solution arises. However,
 before doing that we are going to derive the fuzzy 2-sphere solution of
 \cite{BMN} in our notation, and we are going to compare it with the calculation in terms of a 
 classical spherical M2-brane that we have included in the Appendix for the sake
 of completeness.
 
 The BMN Matrix model contains a fuzzy 2-sphere solution with non-zero momentum $p_-$
and zero light-cone energy, which is the microscopical realization of the classical
2-sphere giant graviton of \cite{GST} in the pp-wave background \cite{BMN}.
Taking the ansatz $r={\rm const}$, $y=Z^a=0$, $a=1,\dots 8$, and (\ref{ansatz1}) for
$(X_1,X_2,X_3)$
in (\ref{newaction}),
we arrive at the following Hamiltonian
 \begin{equation}
 \label{HM2}
 H=\int dx^+ \frac{N}{R}r^2
 \Bigl(\frac{\mu}{6}-\frac{2 Rr}{\sqrt{N^2-1}}\Bigr)^2\, .
 \end{equation}
 Minimizing with respect to $r$ one finds a zero energy solution for $r=0$, which corresponds
 to the point-like graviton, and another
 one for finite $r$:
 \begin{equation}
 \label{radioM2}
 r=\frac{\mu}{12}\frac{\sqrt{N^2-1}}{R}\, ,
 \end{equation}
 which corresponds to the giant graviton solution. When the number of gravitons is large 
 this value of the radius agrees
 with the radius of the classical 2-sphere giant graviton solution in this
background, given by expression (\ref{radiomac}) in the Appendix
 (recall that $T_2=(2\pi)^{-1}$ in our units,
 in which $T_0$, the tension of a single graviton, is equal to 1, and that we
 are describing the sector of the theory with 
 $p_-=-N/R$). 
 Moreover, there is perfect agreement between the corresponding Hamiltonians, 
 given by (\ref{HM2})
 and (\ref{Hmac}),  in this limit.
 
 In the next section we will show that the extra dielectric term present in our action
 (\ref{newaction}) is responsible for the existence of a second zero energy solution
 which is interpreted as the fuzzy 5-sphere giant graviton.
 
 \section{The fuzzy 5-sphere giant graviton}
 
It has been shown in \cite{JLR3} that the 5-sphere giant graviton solutions in 
$AdS_4\times S^7$ and
$AdS_7\times S^4$ are realized microscopically in terms of fuzzy 5-spheres defined as
$S^1$ bundles over fuzzy $CP^2$. 

The fuzzy $CP^2$ has
been extensively studied in the literature, in different contexts 
\cite{ABIY}\cite{NR}-\cite{ABS}. 
Embedding the $CP^2$ in $\R^8$ as in section 3 the simplest way to
construct a fuzzy $CP^2$ is by making the $z_a$ coordinates
non-commutative and impose the constraints (\ref{constr}) at the level of
matrices.  One sees that taking
\begin{equation}
Z^a=\frac{1}{\sqrt{C_N}}T^a
\end{equation}
with $T^a$ the generators of $SU(3)$ in an $N$ dimensional representation, and $C_N$
the quadratic Casimir in this representation, the constraints are satisfied iff the $Z^a$
are taken in the $(n,0)$ or $(0,n)$ representations of $SU(3)$ \cite{ABIY},
parametrizing the
irreducible representations of $SU(3)$ by two integers $(n,m)$ corresponding to the
number of fundamental and anti-fundamental indices (see \cite{JLR3} and references
therein for more details). In these representations
\begin{equation}
\label{rep1}
Z^a=\frac{1}{\sqrt{\frac13 n^2+n}}T^a\, ,
\end{equation}
and the commutation relations of the $Z^a$ become
\begin{equation}
\label{rep2}
[Z^a,Z^b]=\frac{i}{\sqrt{\frac13 n^2+n}}f^{abc} Z^c\, .
\end{equation}
 
 The fuzzy 5-sphere solution emerges when, on top of (\ref{rep1}),
 $r=X^i=0$,  $i=1,2,3$, and $y$ and $Z^a$ are taken time independent in the Hamiltonian
 (\ref{newaction}). Then one gets
 \begin{equation}
 \label{HM5}
 H=\int dx^+ \frac{N}{R}y^2\Bigl( \frac{\mu}{12}-\frac{Ry^4}{4(n^2+3n)}
 \Bigr)^2\, ,
 \end{equation}
 where we have used that
 \begin{equation}
 f_{[abe}f_{cd]f}Z^d Z^c Z^b Z^a Z^e Z^f=-\frac{1}{4(n^2+3n)} \unity\, .
 \end{equation}
 
 Minimizing the Hamiltonian (\ref{HM5}) one finds two zero light-cone energy solutions: 
 one for $y=0$, 
the point-like graviton, and another
 one for
 \begin{equation}
 \label{radioM5}
 y=(\frac{\mu(n^2+3n)}{3R})^{1/4}\, ,
 \end{equation}
 which corresponds to the giant graviton solution. 
 Taking into account that the dimension of the $(n,0)$ and $(0,n)$
 representations is given by
 \begin{equation}
 N=\frac{(n+1)(n+2)}{2}
 \end{equation}
 and that $p_-=-N/R$,
 we have that for large $N$
 \begin{equation}
 y\sim (-\frac23 \mu p_-)^{1/4}
 \end{equation}
 which is the radius of the classical 5-sphere solution discussed in the Appendix,
 where we have to take into account that $T_5=(8\pi^3)^{-1}$ in our units. Also, in this
 limit, the Hamiltonians (\ref{HM5}) and (\ref{Hmac}) agree exactly.
 This agreement is an important check for the validity of our Matrix model
 (\ref{newaction}).
 
\section{Conclusions}

Using the action for $N$ coincident gravitational waves in M-theory we have
constructed a pp-wave Matrix model containing the fuzzy 5-sphere giant graviton
as a supersymmetry preserving solution. Our Matrix model is the sum of the
(bosonic part of the) BMN Matrix action\footnote{With the remarks that we have made
after equation (\ref{newaction}).}
plus an additional quadrupolar coupling
to the 6-form potential of the pp-wave background. This coupling is the one
responsible for the expansion of the waves into a fuzzy 5-sphere. 
We have seen that in order to see this new coupling arising it is necessary to
go to a more non-perturbative regime in M-theory. In this regime the waves 
are related to Type IIA waves upon reduction.

We should emphasize that we have only worked out the bosonic terms in the action.
However the agreement between our fuzzy 2- and 5-sphere solutions and the
classical description in \cite{GST} suggests that both should occur as BPS
solutions of a supersymmetric Matrix action,
preserving the same half of the supersymmetries as the point-like
graviton \cite{GMT}.  It would be interesting to see if this
Matrix action could be derived as in
\cite{BMN} keeping the 6-form potential of
the background and using the supervielbeins for $AdS_4\times S^7$ and
$AdS_7\times S^4$ of \cite{WPPS}.

A Matrix action having a fuzzy 5-sphere giant graviton solution has also been 
given recently in \cite{Nastase}. The approach taken in this reference is 
to construct the Matrix
action such that the defining algebra of the fuzzy 5-sphere of \cite{Ram2} 
is obtained from its
equations of motion. The fuzzy 5-sphere defined in \cite{Ram2} is however
different from the fuzzy 5-sphere that we have obtained in this paper, as solution to
our Matrix model. One obvious difference is that the fuzzy $S^5$ in \cite{Ram2}
has explicit $SO(6)$ invariance whereas our construction is only $SU(3)\times U(1)$
manifestly invariant\footnote{The $SO(6)$ invariance might still be present in a 
non-manifest way, in the same way the classical $S^5$ is not explicitly $SO(6)$
invariant when it is described as an $S^1$ bundle over $CP^2$.}.
Another difference is that our solution approaches neatly the
classical $S^5$ in the large $N$ limit, where all the non-commutativity disappears,
whereas this is not the case for the fuzzy 5-sphere of \cite{Ram2}. Other more
technical differences can be found in \cite{JLR3}. Therefore there are clear differences
between both fuzzy sphere constructions. From this point of view one should not
expect exact coincidence between the Matrix model presented in \cite{Nastase} and
our Matrix model. 

We believe one nice feature about our Matrix model is that it reproduces the right value
for the radius of the 5-sphere giant graviton solution. This value cannot however be
predicted from
the Matrix model in \cite{Nastase}, because the requirement of reproducing the
defining algebra of the fuzzy 5-sphere of \cite{Ram2}
from its equations of motion is not enough to
fix completely all the parameters in the action.
Moreover, some arguments
suggest that the radius of this giant graviton solution should scale with the light-cone
momentum as $r\sim (p_+)^{1/5}$ \cite{Ram1,S-J}, so it would not reproduce the correct
value for the classical 5-sphere found in \cite{GST,MSR}  in the large $N$ limit. 
The description of the
fuzzy 5-sphere solution as an $S^1$ bundle over a fuzzy $CP^2$ seems to be an
essential ingredient towards the correct increasing of the power of the radius 
of the solution with the light-cone momentum.

Our Matrix model arises from the action for coincident M-theory gravitational waves
constructed in \cite{JL2}. Using dualities similar actions have been derived
for Type II gravitational waves (see \cite{JL1}, \cite{JLR}). For consistency the pp-wave
Matrix models, constructed in the literature using different approaches,
in Type IIA \cite{B,SY2,HS,DMS}  and 
Type IIB \cite{S-J} 
should also arise from these actions when the waves propagate in the pp-wave
 background
with non-vanishing light-cone momentum. From this point of view
the physical
interpretation of the fundamental constituents referred as 
{\it tiny gravitons} in references \cite{S-J,ST} would be as 
gravitational waves, and, clearly, the theory describing them would be a 
$U(N)$ gauge theory, since these actions arise through dualities from the $U(N)$
gauge theory describing M-theory gravitational waves. These dualities imply 
in particular that
the vector field in the Type IIA Matrix model would be associated to D2-branes
wrapped on the direction of propagation of the waves, ending on them, 
and in the Type IIB Matrix model would be associated to D3-branes wrapped on the
direction of propagation and also on a second compact direction (the
T-duality direction from Type IIA) ending on them. We hope to report the details of these
calculations in a forthcoming paper \cite{LR}.

\subsubsection*{Acknowledgements}

It is a pleasure to thank Bert Janssen for useful discussions.
This work has been
partially supported by CICYT grant BFM2003-00313 (Spain), and by the European
Commission FP6 program MRTN-CT-2004-005104, in which the authors are associated
to Universidad Aut\'onoma de Madrid.
D.R-G. was supported in part by a F.P.U.
Fellowship from M.E.C. (Spain).

\appendix 
\renewcommand{\theequation}{\Alph{section}.\arabic{equation}}
\sect{Macroscopical giant gravitons for maximally supersymmetric pp-wave backgrounds}

In this appendix we summarize the macroscopical description of the giant graviton
solutions for the pp-wave backgrounds that arise as Penrose limits of 
$AdS_m\times S^n$ (see \cite{MSR,DMS,SS}). 

The Penrose limits of the $AdS_4\times S^7$ and 
$AdS_7\times S^4$ spacetimes give rise to the background (\ref{ppwave}), whereas that of
$AdS_5\times S^5$ gives rise to \cite{BFHP2}:
\begin{eqnarray}
\label{ppwave2}
&&ds^2=-4dx^+dx^--\mu^2(x_1^2+\dots +x_8^2)(dx^+)^2+d\vec{x}^2\, ,
\qquad {\rm where} \,\, \vec{x}\in\R^8\nonumber\\
\nonumber\\
&&C^{(4)}_{+\alpha_1\alpha_2\alpha_3}=-\mu r^4 \sqrt{g_\alpha}\, ,\qquad
C^{(4)}_{+\gamma_1\gamma_2\gamma_3}=\mu y^4\sqrt{g_\gamma}\, ,
\end{eqnarray}
where $\{\alpha_i\}$ and $\{\gamma_i\}$ are the angle variables parametrizing
the two 3-spheres embedded in $\R^8$, and $r^2=x_1^2+\dots +x_4^2$,
$y^2=x_5^2+\dots +x_8^2$.

The giant graviton solutions associated to these backgrounds are
spherical 2-branes and 5-branes  
for the Penrose limit of $AdS_4\times S^7$ and $AdS_7\times S^4$,
and spherical 3-branes for the background (\ref{ppwave2}). In all
cases the branes are stable thanks to their dipole or magnetic moment with 
respect to the background potential.

Taking the ansatz for a $p$-brane giant graviton solution in these backgrounds 
one has that, generically
\begin{equation}
ds^2=-4dx^+ dx^- -\beta_p (dx^+)^2+d\vec{x}^2
\end{equation}
where $\vec{x}$ parametrizes a point in $\R^{p+1}$,
$\beta_p=\sigma_p {\mu}^2 r^2$ with $\sigma_2=\frac19$, $\sigma_3=1$
and $\sigma_5=\frac{1}{36}$, and $r$ is the radius of the $p$-brane,
$r^2=x_1^2+\dots +x_{p+1}^2$. The non-vanishing background potential can also
be written in a unified way, as
\begin{equation}
C^{(p+1)}_{+\alpha_1\dots \alpha_p}=-\sqrt{\sigma_p}\mu r^{p+1}\sqrt{g_\alpha}\, .
\end{equation}
We fix a gauge in which the spatial
coordinates are identified with the angular coordinates and $t=x^+$. The $p$-brane
moves along the $x^-$ direction by taking $x^-=x^-(x^+)$.

Substituting this trial solution into the worldvolume action of the $p$-brane and
integrating the angular worldvolume coordinates
one finds
\begin{equation}
S_p=T_{p} A_{p} \int dx^+ \Bigl\{ -r^{p}\sqrt{\sigma_p \mu^2 r^2+4\dot{x}^-}
+\sqrt{\sigma_p}\mu r^{p+1}\Bigr\}
\end{equation}
where $A_{p}$ is the area of a unit $p$-sphere.

Legendre transforming $\dot{x}^-$ to $p_-$ one finally arrives at a Hamiltonian
\begin{equation}
\label{Hmac}
H=-\int dx^+ p_- r^2 \Bigl( \frac12 \sqrt{\sigma_p} \mu+\frac{T_p A_p r^{p-1}}
{p_-}\Bigr)^2
\end{equation}
from which one finds that the light-cone energy vanishes for $r=0$, which
corresponds to the point-like graviton, and for
\begin{equation}
\label{radiomac}
r=(-\frac{\sqrt{\sigma_p}\mu p_-}{2T_{p}A_{p}})^{1/(p-1)}
\end{equation} 
which corresponds to the giant graviton solution. 


\end{document}